# Precise estimation of the *S* = 2 Haldane gap by numerical diagonalization

Hiroki Nakano[1] and Toru Sakai,[1,2]

[1]*Graduate School of Material Science, University of Hyogo, Kamigori, Hyogo 678-1297, Japan*

[2]*National Institutes for Quantum and Radiological Science and Technology (QST), SPring-8 Sayo, Hyogo 679-5148, Japan*

The Haldane gap of the *S*=2 Heisenberg antiferromagnet in a one-dimensional linear chain is examined by a numerical-diagonalization method. A precise estimate for the magnitude of the gap is successfully obtained by a multistep convergence-acceleration procedure applied to finite-size diagonalization data under the twisted boundary condition.

It is well-known that the Haldane conjecture[1, 2] for gapped one-dimensional Heisenberg antiferromagnets for integer *S* provides considerable contributions to our understanding of quantum spin systems. The energy gap in the spin excitation – the Haldane gap – has been extensively studied. The magnitude of the gap has been estimated by various approaches. The Haldane gap is the largest for *S*=1; the density matrix renormalization group (DMRG),[3] quantum Monte Carlo (QMC),[4] and numerical-diagonalization (ND)[5] calculations give estimates that agree with each other to a very precise number of digits: $\Delta/J \sim 0.41048$, where $\Delta$ and *J* denote the gap value and the strength of the interaction defined later, respectively.

Let us observe the situation for *S*=2, for which there are experimental studies exploring materials as its candidates.[6, 7] Since the Haldane gap for *S*=2 is smaller than that for *S*=1, even though various theoretical studies have attempted to improve the estimate, the precision remained low until Wang et al.[8] concluded from their DMRG study that the gap is estimated as

$$\Delta/J = 0.0876 \pm 0.0013. \qquad (1)$$

On the other hand, Todo and Kato[4] carried out QMC simulations and obtained

$$\Delta/J = 0.08917 \pm 0.00004. \qquad (2)$$

Note here that these estimates in (1) and (2) are close but different from each other within their own errors. The difference is very small but should be resolved as fundamental information in condensed-matter physics. Ueda and Kusakabe[9] carried out DMRG calculations applied to systems with tuned interactions at both ends of the chain and concluded that

$$\Delta/J = 0.0891623 \pm 0.0000009. \tag{3}$$

The estimate in (3) agrees with that in (2) but disagrees with that in (1) within their own errors. The disagreement between the estimates in (1) and (3) suggests that extrapolation from DMRG calculations is quite sensitive to how the boundary effects are treated and that a highly careful treatment is required.

Next, let us observe the situation of ND studies of the $S=2$ Haldane gap. It is well-known that the ND method can treat very small clusters because the dimension of the Hilbert space grows exponentially with respect to the system sizes. This is the reason why the ND method has been almost powerless for the estimation of the $S=2$ Haldane gap until the twisted boundary condition (TBC) was found to be very useful for estimation.[5] In ref. 5, a convergence-acceleration technique is applied to a sequence of finite-size gaps up to 16 sites under the TBC; the gap is concluded to be

$$\Delta/J = 0.0886 \pm 0.0018. \tag{4}$$

Since the estimate in (4) includes a large error, it agrees with the estimates in (1), (2), and (3). The reason why the error in the estimate in (4) is large is that the treated system sizes are up to 16 spin sites in ref. 5. The smallness of the maximum of the sizes has a negative influence that the convergence-acceleration procedure can be carried out only once but cannot be repeated for $S=2$, while the procedure can be repeated four times for $S=1$.

Under these circumstances, the present paper has two purposes. The primary purpose is to present new information concerning the issue of two disagreeing estimates of the $S=2$ Haldane gap from the viewpoint of the ND method that is unbiased against any approximation. Note here that this method can give some information neutrally because it is different from the QMC and DMRG methods. To do it successfully, we should carry out the ND calculations for even larger finite-size clusters than those reported in ref. 5. If such additional data are obtained, one can examine whether or not the convergence-acceleration procedure can be repeated. This examination is the secondary purpose of this paper.

The Hamiltonian that we study is given by $H = \sum_i J S_i \cdot S_{i+1}$, where $S_i$ denotes the $S=2$ spin operator at site $i$. The label of a spin site is denoted by $i$, which is assumed to be an integer; the number of spin sites is denoted by $N$. Energies are measured in units of $J$; hereafter, we set $J=1$. Numerical diagonalization is carried out in the subspace belonging to $\sum_j S^z_j = M$. In this paper, we investigate the lowest eigenvalues of the isotropic system in the spin space; therefore, our diagonalization is carried out for $M=0$. Numerical diagonalization is based on the Lanczos algorithm, which provides us with the lowest energies in the target subspace. Part of the Lanczos diagonalizations were carried out using an MPI-parallelized code, which was originally developed in ref. 5. The usefulness of our program was confirmed with large-scale parallelized calculations.[10–17] In the present study, the maximum system size is $N=20$, which is a new world record to the best of our knowledge. The dimension of the matrix of the subspace $M=0$ for $N=20$ is 5,966,636,799,745, which is larger than 712,070,156,203 of the 27-site $S=1$ case in which Lanczos diagonalization was successfully carried out.[11] We impose the TBC that is given by $S^x_{N+1} = -S^x_1$, $S^y_{N+1} = -S^y_1$, $S^z_{N+1} = S^z_1$. As mentioned above, this boundary condition can help us to estimate Haldane gaps even when the integer $S$ is large.

In this paper, we attempt to extrapolate the sequences of the ND raw data for the finite-size clusters by convergence acceleration. We apply Wynn's epsilon algorithm[18] given by $[A^{(k+1)}_N - A^{(k)}_{N-2}]^{-1} = [A^{(k)}_{N-4} - A^{(k)}_{N-2}]^{-1} + [A^{(k)}_N - A^{(k)}_{N-2}]^{-1} - [A^{(k-1)}_{N-4} - A^{(k)}_{N-2}]^{-1}$ when $A^{(-1)}_N = \infty$. This algorithm was successfully used to estimate the $S=1$ Haldane gap.[5,19] Here, we will substitute ND data determined by $E_0$ and $E_1$ into the initial sequence $A^{(0)}_N$, where $E_0$ and $E_1$ are the ground-state and first-excited-state energies for a given $N$, respectively. We focus our attention on the first excitation gap, namely, $G(N) \equiv E_1 - E_0$. It is convenient to estimate the decay lengths defined as $\xi^{(k)}_N = 2/\log([A^{(k)}_{N-4} - A^{(k)}_{N-2}]/[A^{(k)}_{N-2} - A^{(k)}_N])$ in order to examine the convergence behavior of $A^{(k)}_N$ for each step $k$. When $A^{(k)}_N$ for each $k$ is monotonic with respect to $N$, a monotonic increase in $\xi^{(k)}_N$ suggests that the convergence of $A^{(k)}_N$ with respect to $N$ becomes slower. The confirmation of whether or not the condition $\xi^{(k+1)}_N < \xi^{(k)}_N$ holds enables us to find whether the acceleration of $A^{(k)}_N$ from $A^{(k-1)}_N$ is successful or unsuccessful for each $k$. If it is successful, a sequence $B^{(k)}_{N+1}$ defined as $[T_N S_{N+2} - T_{N+2} S_N]/[S_{N+2} - S_N - T_{N+2} + T_N]$ is convergent from the side opposite to $A^{(k)}_N$, where $T_N = A^{(k)}_N$ and $S_N = A^{(k-1)}_N$. From $A^{(k)}_N$ and $B^{(k)}_N$, we can obtain the extrapolated estimate as well as its error.

Now, let us examine $G(N)$; the results are presented in Table I. Note here that results up to $N = 16$ were reported in ref. 5 and that our results for $N = 18$ and 20 are additionally presented in this paper. One finds that the data for small $N$ disturb the table. In Table I, $\xi^{(2)}_{16}$ is negative. This behavior originates from $A^{(0)}_4$. This disturbance was reported in ref. 5. In the present study, one additionally finds that $\xi^{(2)}_{18} > \xi^{(1)}_{18}$. This behavior suggests that it is unclear whether or not the acceleration of convergence is successful. This behavior originates from $A^{(0)}_6$. In Table I, the data originating from $A^{(0)}_4$ and $A^{(0)}_6$ are underlined. When we exclude these underlined data from the table, all of the other data do not disturb the table, which suggests that the convergence acceleration of the sequence $A^{(0)}_N$ starting from $N = 8$ is successful. We can thus confirm that $A^{(2)}_{16}$, $A^{(2)}_{18}$, and $A^{(2)}_{20}$ are successfully accelerated. We can then employ $A^{(2)}_{20}$ as a lower bound. One can also find that $B^{(2)}_N$ without the underline shows a dependence opposite to $A^{(k)}_N$. It is also confirmed that $B^{(2)}_N$ approaches its limit from the larger side even after we exclude the underlined data. Thus, we can employ $B^{(2)}_{19}$ as an upper bound. For the above reason, it is a careful and reliable judgment to consider that $\Delta(S = 2)$ is between $A^{(2)}_{20}$ and $B^{(2)}_{19}$. Our conclusion for the first excitation gap is

$$\Delta/J = 0.0890 \pm 0.0007, \qquad (5)$$

assuming the monotonic behaviors of $A^{(2)}_N$ and $B^{(2)}_N$ for $N > 20$. This estimate with its error completely includes the QMC estimate in (2) and the DMRG estimate in (3). On the other hand, the central value of the estimate in (1) is not included within the error of the present estimate in (5). Although the two estimates in (1) and (5) share part of their errors, the present study suggests that the QMC estimate in (2) and the DMRG estimate in (3) are more reliable than the estimate in (1).

In summary, we have demonstrated a successful multistep convergence-acceleration procedure of ND data to estimate the Haldane gap of the $S=2$ Heisenberg antiferromagnetic chain precisely. If the ND results for systems as large as possible are appropriately obtained, one can find a reliable and precise estimate of a physical quantity, which greatly contributes to our understanding of quantum spin systems.

**Acknowledgements**

We wish to thank Professors N. Todoroki and H. Tadano for fruitful discussions. This work was partly supported by JSPS KAKENHI, grant numbers 16K05418, 16K05419, 16H01080 (JPhysics), and 18H04330 (JPhysics). In this research, we used the computational resources of the K computer provided by the RIKEN Advanced Institute for Computational Science through HPCI System Research projects (Project IDs: hp170018, hp170028, and hp170070). We used the computational resources of Fujitsu PRIMERGY CX600M1/CX1640M1 (Oakforest-PACS) provided by the Joint Center for Advanced High Performance Computing (JCAHPC) through HPCI System Research projects (Project IDs: hp170207 and hp180053). The computational resources of Oakforest-PACS were also awarded by the "Large-scale HPC Challenge" Project of JCAHPC. Some of the computations were performed using the facilities of the Department of Simulation Science, National Institute for Fusion Science; the Institute for Solid State Physics, The University of Tokyo; and the Supercomputing Division, Information Technology Center, The University of Tokyo.

Table I. Sequence of finite-size gaps for the $S = 2$ case under the twisted boundary condition, the convergence acceleration based on Wynn's algorithm, and the upper-bound sequence $B^{(k)}_N$. The ground-state energies per site for each $N$ are also presented.

| $N$ | $-E_0/N$ | $10^2 A_N^{(0)}$ | $\xi_N^{(0)}$ | $10^2 A_N^{(1)}$ | $\xi_N^{(1)}$ | $10^2 B_N^{(1)}$ | $10^2 A_N^{(2)}$ | $\xi_N^{(2)}$ | $10^2 B_N^{(2)}$ |
|---|---|---|---|---|---|---|---|---|---|
| 4 | 4.611858407606 | 3.60543243815 | | | | | | | |
| 6 | 4.699792263067 | 5.88636574630 | | | | | | | |
| 8 | 4.728484454192 | 6.98099292140 | 2.72 | 7.9910262 | | | | | |
| 9 | | | | | | 9.046111 | | | |
| 10 | 4.741217388690 | 7.57841496301 | 3.30 | 8.2962537 | | | | | |
| 11 | | | | | | 9.125346 | | | |
| 12 | 4.747911544485 | 7.93926067768 | 3.97 | 8.4896523 | 4.38 | | 8.6729401 | | |
| 13 | | | | | | 9.068141 | | | 8.814100 |
| 14 | 4.751834689354 | 8.17386831221 | 4.65 | 8.6098758 | 4.21 | | 8.7252465 | | |
| 15 | | | | | | 9.021915 | | | 8.963622 |
| 16 | 4.754315127036 | 8.33497991928 | 5.32 | 8.6881548 | 4.66 | | 8.7779957 | -237.32 | |
| 17 | | | | | | 8.991283 | | | 8.968069 |
| 18 | 4.755973933223 | 8.45035858794 | 5.99 | 8.7414450 | 5.20 | | 8.8141819 | 5.306 | |
| 19 | | | | | | 8.971044 | | | 8.955061 |
| 20 | 4.757132281620 | 8.53576315931 | 6.65 | 8.7791046 | 5.76 | | 8.8390183 | 5.314 | |